\documentclass{article}

\usepackage{PRIMEarxiv}

\usepackage[utf8]{inputenc} 
\usepackage[T1]{fontenc}    
\usepackage{hyperref}       
\usepackage{url}            
\usepackage{booktabs}       
\usepackage{amsfonts}       
\usepackage{nicefrac}       
\usepackage{microtype}      
\usepackage{lipsum}
\usepackage{fancyhdr}       
\usepackage{graphicx}       
\usepackage{amsmath}
\usepackage{amssymb}
\usepackage{booktabs}
\usepackage{subcaption}
\usepackage{xcolor}
\usepackage{algorithm}
\usepackage{algorithmic}
\graphicspath{{media/}}     

\newcommand{\methname}{HetACUMN}

\newcommand{\myeqref}[1]{Eq. \eqref{#1}}
\newcommand{\cref}[1]{\ref{#1}}

\definecolor{darkgreen}{RGB}{1, 50, 32}
\definecolor{limegreen}{RGB}{50, 205, 50}

\title{Improved Cryo-EM Pose Estimation and 3D Classification through Latent-Space Disentanglement}

\author{
Weijie Chen\textsuperscript{a,b},
Yuhang Wang\textsuperscript{a*},
and
Lin Yao\textsuperscript{a*}\thanks{Corresponding authors}\\
\textsuperscript{a} DP Technology, Ltd., Beijing, China;
\textsuperscript{b} Peking University, Beijing, China\\
\texttt{baoz@pku.edu.cn, stevenwaura@gmail.com, yaol@dp.tech
}
}

\begin{document}
\maketitle

\begin{abstract}
Due to the extremely low signal-to-noise ratio (SNR) and unknown poses (projection angles and image shifts) in cryo-electron microscopy (cryo-EM) experiments, reconstructing 3D volumes from 2D images is very challenging. In addition to these challenges, heterogeneous cryo-EM reconstruction requires conformational classification. In popular cryo-EM reconstruction algorithms, poses and conformation classification labels must be predicted for every input cryo-EM image, which can be computationally costly for large datasets. An emerging class of methods adopted the amortized inference approach. In these methods, only a subset of the input dataset is needed to train neural networks for the estimation of poses and conformations. Once trained, these neural networks can make pose/conformation predictions and 3D reconstructions at low cost for the entire dataset during inference. Unfortunately, when facing heterogeneous reconstruction tasks, it is hard for current amortized-inference-based methods to effectively estimate the conformational distribution and poses from entangled latent variables. Here, we propose a self-supervised variational autoencoder architecture called ``\methname{}'' based on amortized inference. We employed an auxiliary conditional pose prediction task by inverting the order of encoder-decoder to explicitly enforce the disentanglement of conformation and pose predictions. Results on simulated datasets show that \methname{} generated more accurate conformational classifications than other amortized or non-amortized methods. Furthermore, we show that \methname{} is capable of performing heterogeneous 3D reconstructions of a real experimental dataset.
\end{abstract}

\section{Introduction}
Cryo-electron microscopy (cryo-EM) is an experimental method
for resolving the structures of 3-dimensional (3D) objects
(typically biological molecules) from 2-dimensional (2D)
projection images.
This technique
has seen rapid progress in the past decade.
The 3D structures of many important biological molecules have been
resolved based on this method, such as the SARS-Cov-2
relevant proteins~\cite{Ye_2022_SARS}.
Recognizing its importance, the 2017 Nobel prize in chemistry
was awarded to the three pioneers of cryo-EM~\cite{Nobel_Chemistry_2017}.

There are three challenging aspects of cryo-EM reconstructions that
are different from 3D reconstructions from natural images
commonly seen in computer vision area.
First, the signal-to-noise ratio (SNR) of cryo-EM images is
very low (around -10 dB), making it hard to distinguish signals
from noise. The distribution of noise is also complex. It can vary significantly between various datasets due to different experimental conditions and the nature of target sample objects.
The second challenge is to recover the projection angle and
2D image translations (collectively termed ``poses'')
from images alone. The quality of final reconstructed objects depends the accuracy of the estimation of pose.
In single-particle cryo-EM,
the pose information is completely missing during experimental
measurements, and can only be estimated based on a large set
of projections.
On top of these two challenges,
the cryo-EM 3D reconstruction problem is further compounded by
the fact that the underlying objects exist in many possible
conformations.
Here the conformations refer to
various possible shapes of the same molecule (e.g., a protein), akin to different postures of a human body.
Heterogeneous cryo-EM 3D reconstruction is concerned with solving
all three challenges, which is a difficult and heavily researched direction
~\cite{Donnat_JSB_2022}.

Recent advances in deep learning and computer vision provided new
paths for solving these challenges, as have been demonstrated in
previous work such as cryoDRGN2~\cite{Zhong_2021_ICCV} and cryoFIRE~\cite{Levy_2022_cryoFIRE}.
CryoDRGN2 can generate accurate classifications and reconstructions of heterogeneous EM datasets~\cite{Zhong_2021_ICCV}, but the computational cost is high for large datasets~\cite{Levy_2022_cryoFIRE}.
CryoFIRE demonstrated significant speed-up compared to CryoDRGN2, but at the cost of reduced accuracy~\cite{Levy_2022_cryoFIRE}.
To overcome these problems, here we propose a new method called~\methname{} (Heterogeneous reconstruction with Amortized Classification Using
Multi-task Networks)
with the following highlights.
\vspace*{0pt}
\begin{itemize}
    \item We propose a novel training framework for enforcing disentanglement of pose and conformation, which is specifically designed for heterogeneous reconstructions.
    \item We demonstrate the effectiveness of our method on simulated datasets, achieving high-quality reconstructions and better performance of conformational classification.
    \item Our method can also be applied for heterogeneous reconstruction of experimental datasets, demonstrating its value in protein structure studies.

\end{itemize}

\section{Background}
In cryo-EM, a large number of macromolecules are frozen with random orientation and independent conformations in a thin layer of vitreous ice. Probing electrons interact with
electrostatic potential created by these molecules and produce two-dimensional tomographic projection images.
The volumes $\{V_i\}_{i=1,...,N}$ can be assumed as a subset of conformations independently sampled from the
conformational space of macromolecules. It is assumed that conformational space $\mathcal{M}$ can be characterized by a
low-dimension manifold and embedded in $\mathbb{R}^{d}$.

On the screen of observer, each molecule is in unknown orientation $R_i \in SO(3) \subset \mathbb{R}^{3\times3}$
and conformational state $z_i \in \mathbb{R}^{d}$. The electron beam interacts with the molecules and projects a 2D image $\mathcal{I}_i$.
Besides, the system's impulse response can be represented by the Point Spread Function (PSF) $P_i$, which can be estimated
based on microscope imaging parameters.
More specifically, each image can be modeled as follows:
\begin{equation}
  \mathcal{I}_i(x,y) = T_{t_i}*P_i*\int_{\omega}V(R_i\cdot[x,y,\omega]^{T}, z_i)d\omega + \epsilon_i
  \label{eq:real_image}
\end{equation}
where * is the convolution operation, $T_{t}$ is the translation operator characterized by $t \in \mathbb{R}^{2}$, $\epsilon_i$ is noise.

Based on the Fourier slice theorem~\cite{doi:10.1146/annurev-biodatasci-021020-093826}, we can bypass the computation of integrals and convolutions in \myeqref{eq:real_image}.
\begin{equation}
  \mathcal{F}_{2D}[\int_{\omega}V(R_i^{T}\cdot[x,y,\omega], z_i)d\omega] = \mathcal{F}_{3D}[V](R_i^{T}\cdot[k_x, k_y, 0], z_i)
\end{equation}
 where $\mathcal{F}_{2D}$ and $\mathcal{F}_{3D}$ denote 2D and 3D Fourier transform respectively. Finally, \myeqref{eq:real_image} can be rewritten with
 Fourier transform $\hat{\mathcal{I}}=\mathcal{F}_{2D}[\mathcal{I}]$ and $\hat{V}=\mathcal{F}_{3D}[V]$:
 \begin{equation}
  \hat{\mathcal{I}}_i(k_x, k_y) = \hat{T}_{t_i}\odot C_i \odot \hat{V}(R_i^{T}\cdot[k_x, k_y, 0], z_i) + \hat{\epsilon}_i
  \label{eq:fourier_image}
\end{equation}
where $\odot$ is element-wise multiplication, $C_i=\mathcal{F}_{2D}[P_i]$ is the
Contrast Transfer Function (CTF)~\cite{marabini2016electron}, $\hat{T}_{t}$ is the translation operator in Fourier space.

\section{Related Work}
Here we review existing methods relevant to heterogeneous cryo-EM reconstructions.

\subsection{Conformational classification}
In traditional cryo-EM software such as RELION~\cite{Scheres_2012_JSB} and cryoSPARC~\cite{2017cryoSPARC},
conformational heterogeneity is usually modeled as a set of discrete
conformations for simplicity
using expectation-maximization-based algorithms.
Alternatively, the heterogeneity
can be represented continuously.
One class of such methods use deformation-based approaches,
such as rigid bodies\cite{scheres2012relion},
normal-mode-based eigenvolumes~\cite{10.3389/fmolb.2021.663121},
3D linear subspace models~\cite{Punjani_2021_3DVA},
deformation field
~\cite{Herreros_2021_deformation_field,Punjani_2022_3DFlex}
and high-dimensional molecules (``hyper-molecules'')~\cite{Lederman_2020_hypermol}.
Another class uses a neural-network-based representation
to encode conformational variability implicitly into network parameters,
which include
cryoDRGN-BNB~\cite{Zhong_2019_cryodrgn_bnb}, cryoDRGN2~\cite{Zhong_2021_ICCV}, and cryoFIRE~\cite{Levy_2022_cryoFIRE}.
This line of approach doesn't require any prior knowledge or initial
guesses of the conformational distribution as used in the deformation-based
methods, and is therefore less biased.
\methname{} also adopted this approach.

\begin{figure*}[!th]
  \begin{center}
    \includegraphics[width=0.95\linewidth]{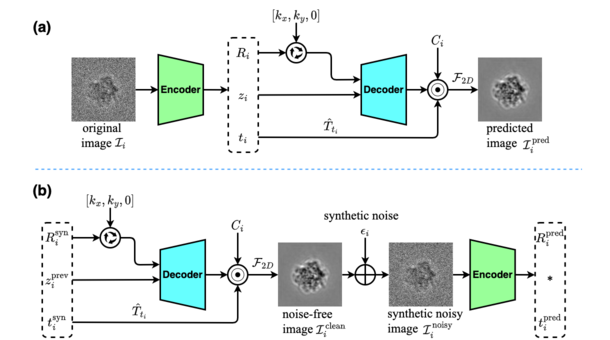}
  \end{center}
     \caption{Architectures of the two tasks in \methname{}. (\textbf{a}) the variational image reconstruction task; (\textbf{b}) the conditional pose prediction (\textbf{CPP}) task.}
  \label{fig:ipi}
\end{figure*}

\subsection{Pose estimation}
Pose estimation refers to estimating projection angles and image
translations (collectively termed as ``poses'') only from
2-dimensional (2D) cryo-EM projection images.
It is one of the hardest problems in cryo-EM, especially in heterogeneous
cryo-EM reconstruction.

In the majority of literature, pose estimation is
performed on a per-image basis under
the expectation-maximization framework, including
RELION\cite{scheres2012relion}, cryoSPARC\cite{2017cryoSPARC}, 3DFlex\cite{Punjani_2022_3DFlex}, cryoDRGN-BNB\cite{Zhong_2019_cryodrgn_bnb}, cryoDRGN2\cite{Zhong_2021_ICCV}, and cryoFold\cite{2019CryoFold}.
Although these methods can usually reach good accuracy, the
associated computational cost is significant for large datasets.
(see~\cite{Donnat_JSB_2022} for an up-to-date in-depth review).
There has also been attempts to bypass the pose estimation problem
through the generative adversial network as implemented in
cryoGAN~\cite{Gupta_2021_IEEE}, but the reconstruction resolution is
inferior to other methods.

To reduce the computational cost while maintaining high reconstruction quality,
recent development
focused on an alternative approach called amortized inference.
In this approach, the pose estimation problem is formulated as
optimizing the parameters $\xi$ of a function $f_\xi(\mathcal{I})=(R, t)$
that maps an input image $\mathcal{I}$ to its associated rotation $R$ and translation $t$.
Earlier attempts of amortized-inference-base methods have limited accuracy,
e.g., FSTDiff~\cite{Ullrich_2019_PMLR} and CryoPoseNet~\cite{Nashed_2021_ICCV}.
CryoAI~\cite{Levy_2022_ECCV} showed that amortized-inference-based methods
can easily get trapped in local minima which leads to spurious planar
symmetry. To overcome this drawback, CryoAI proposed to use symmetrized
loss (see Methods section for details) and reached pose estimation accuracy
on par with non-amortized methods.
ACE-EM~\cite{yao_2023_ace} proposed a dual-task training framework to improve the accuracy of pose
estimation.
Besides the usual self-supervised learning for image reconstruction, they designed an extra supervised learning for pose prediction.

\subsection{Joint estimation of conformations and poses}

AtomVAE~\cite{DBLP:journals/corr/abs-2106-14108} is one of the early attempts to use
amortized-inference-based joint estimation based on the VAE architecture,
which showed promising results on simulated datasets.
It discussed several key observations such as the importance of disentangling the conformational latent space and pose latent space and the pre-training of the pose prediction using the fixed base structure.
More recently, combing cryoDRGN and cryoAI's architectures,
cryoFIRE~\cite{Levy_2022_cryoFIRE}
demonstrated that it is possible to obtain good reconstruction quality using
joint amortized inferences on both simulated and experimental datasets.
In cryoFIRE, input images are fed into a Convolutional Neural Network (CNN) to extract visual features. Then, through three multi-layer perceptrons (MLP), the pose parameters (i.e., orientation $R$ and translation $t$) and image classification labels (i.e., conformation $z$) are estimated from the visual features.
They propose a ``pose-only phase'' to enable the model to converge to a consensus volume. They disable the optimization of the conformation MLP at the early stage of training and sample from a normalized Gaussian distribution. Here, we show that this strategy is not sufficient to disentagling conformational latent space and pose latent space. Inspired by ACE-EM~\cite{yao_2023_ace}, we proposed a conditional supervised pose learning task for the
joint pose-conformation estimation and achieved state-of-the-art performance on different heterogeneous datasets.

\section{Methods}
\label{sec:method}

\subsection{Overview of \methname{}}
\methname{} consists of two tasks (Figure~\ref{fig:ipi}).
One task is for variational image reconstruction.
Raw EM images $\mathcal{I}_i$ are fed into an encoder
for predicting three latent variables:
orientation $R_i$, image shift $t_i$, and conformational state $z_i$.
The decoder acts as a physics-based simulator of cryo-EM image projection to represent the Fourier volume $\hat{V}$. It takes the predicted conformational state and the grid of $L^{2}$ 3D-coordinates rotated by estimated rotation, then outputs the corresponding projection images. With the estimated translation and the given CTF parameters $C_i$, the projection images are post-processed to obtain a noise-free reconstruction of input images.

The second task is for conditional pose prediction (\textbf{CPP}), which exploits the same encoder-decoder as the first task but in reversed order to explore larger pose spaces. Instead of image reconstruction, it reconstructs the corresponding projection poses from randomly sampled poses with the reversed pipeline, and minimizes the difference between the pose pairs, i.e., conditional pose prediction task.

In this method, we can balance the pose estimation and image reconstruction by alternating two tasks to optimize the VAE-based model during training.

\subsection{Variational image reconstruction task}
The variational image reconstruction task follows the standard self-supervised learning with the usual encoder-decoder order.
The main objective of variational image reconstruction task is to minimize the difference between the input images and predicted images. Assuming the image length of $L$, mean squared error (MSE) loss function is often used to measure their gap.
\begin{equation}
  \mathcal{L}_{\text{image}}(\mathcal{I}_i) = \frac{1}{L^{2}}\|\mathcal{I}_i-\mathcal{I}^{\text{pred}}_i\|_{F}^{2}
\end{equation}
where $\|A\|_{F}=\sqrt{\sum_{i,j}|A_{i,j}|^{2}}$ is the Frobenius norm.

To reduce the chance of reconstructing volumes with the spurious
pseudo-mirror symmetry artefact, we followed added a slighted modified version of the
\textit{symmetrized loss} proposed by cryoAI~\cite{Levy_2022_ECCV}:
\begin{equation}
  \mathcal{L}_{\text{sym}} = \min\{\mathcal{L}_{\text{image}}(\mathcal{I}_i), \mathcal{L}_{\text{image}}(\mathcal{R}_h[\mathcal{I}_i])\}
\end{equation}
where $\mathcal{R}_h$ represents horizontal image flipping.

For conformational state, the conformation head of encoder predicts mean $\mu_{z|\mathcal{I}}\in\mathbb{R}^{d}$ and variance $\sigma^{2}_{z|\mathcal{I}}\in\mathbb{R}_{+}^{d}$ for each image, as the variational approximation to the posterior $q(z|\mathcal{I})$. In general, the prior $p(z)$ on the conformational space is a standard Gaussian distribution. We introduce Kullback-Leibler (KL) divergence for optimizing posterior distribution.
\begin{equation}
  \mathcal{L}_{z} = D_{KL}(q(z|\mathcal{I}_i)\| p(z))
\end{equation}

At the early stage of training, we sample the conformational states from the prior distribution and freeze the training of conformation head
to focus on improving pose estimation alone.

Besides, we added an L1-regularization for predicted translation  to prevent unrealistic large shift and keep the reconstructed object near the center of the coordinate grid as much as possible. The total loss of variational image reconstruction task as follows:
\begin{equation}
  \mathcal{L}_{I} = \mathcal{L}_{\text{sym}} + \lambda_z\mathcal{L}_{z} + \lambda_t \frac{1}{2} \|t_{i}^{\text{pred}}\|_{1}
\end{equation}
where $\lambda_z=1\times10^{-4}$, $\lambda_t=1\times10^{-3}$

\subsection{Conditional pose prediction task}
The conditional pose prediction task takes the encoder and decoder from the mulit-class image reconstruction task and reversed order.
Compared to the variational image reconstruction task, the inputs and ground-truth labels are no long EM images, but poses
uniformally sampled over over $SO(3) \times \mathbb{R}^{2}$.
This encoder-decoder order reversal turns the decoder into a
EM image data generator conditioned on
a particular conformation. It also exposes the latent pose variables
for direct supervised learning.
In this task, the conformational states are sampled from the posterior distribution of the previous multi-class image reconstruction task.
The conformational states $z_i^{\text{prev}}$  and sampled poses are fed into the decoder (parameters frozen) to generate noise-free images $\mathcal{I}_{i}^{\text{clean}}$. To generate more realistic inputs for the encoder, the additive noise estimated from the input datasets is added to the predicted noise-free images. Then, the encoder $E_\theta$ takes the synthetic noisy images and predict the poses.

We define the conditional pose estimation loss function as follows, where $\lambda_p=0.1$.
\begin{equation}
  \mathcal{L} = \lambda_p \left[\frac{1}{9}\|R_i^{\text{syn}}-R_i^{\text{pred}}\|_{F}^{2} + \frac{1}{2}\|t_i^{\text{syn}}-t_i^{\text{pred}}\|_{1}\right]
\end{equation}

This task can also be viewed as a data-augmentation technique.
Therefore, the encoder can learn from a comprehensive pose training datasets even if the input EM image dataset is small or has highly biased
pose distribution.

\subsubsection{Noise Generation}
\begin{figure}[!t]
  \begin{center}
    \includegraphics[width=0.4\linewidth]{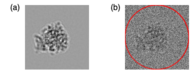}
  \end{center}
     \caption{Example of noisy EM image generation.
     \textbf{(a)} a noise-free reconstructed image.
     \textbf{(b)} a synthetic noisy EM image with SNR = -10 dB. The area outside the red circle is considered as the noise-only region.}
  \label{fig:noise}
\end{figure}

In cryo-EM images, the noise is often simplified as being additive independent, zero-mean Gaussian noise~\cite{scheres2012relion, vulovic2013image}. One can estimate the variance of noise from the corners of the EM images and add synthetic noise to noise-free images directly (Figure~\ref{fig:noise}). However, due to low SNR, this method will overwhelm the detail of images and mislead the encoder in early training.

In our work, we proposed an adaptive noise generation method that generates noise with the fixed SNR instead of using a fixed variance of noise. In other words, the intensity of synthetic noise will increase as the detail of the image reconstruction enhances.
Assuming that the distributions of noise and image signals are independent,  we can estimate the SNR of datasets by following:
  \begin{equation}
  \text{SNR} \triangleq\frac{\sigma_{\mathcal{I}_{\text{clean}}}^{2}}{\sigma_\epsilon^{2}} =  \frac{\sigma_{\mathcal{I}}^{2} - \sigma_\epsilon^{2}}{\sigma_\epsilon^{2}}
  \label{eq:snr}
\end{equation}
where $\sigma_\epsilon^{2}$ is the variance of noise and $\sigma_{\mathcal{I}}^{2}$ is the variance of noisy images.
During training, the noise can be generated from the following Gaussian distribution.
\begin{equation}
  \begin{aligned}
  \epsilon_k \sim& \mathcal{N}(0, \sigma_{\epsilon_k}^{2}) \text{ where }
  \sigma_{\epsilon_k}^{2} = \sigma_{\mathcal{I}_{\text{clean},k}}^{2} / \text{SNR}
  \end{aligned}
\end{equation}
where $\epsilon_k$ is the synthesized noise at the $k$-th step, $\mathcal{I}_{\text{clean},k}$ is the reconstructed noise-freee images at the $k$-th step.

\subsection{Metric for measuring latent-space entanglement}
Currently there are no commonly accepted metrics for measuring
the latent-space entanglement for deep-learning-based heterogeneous 
cryo-EM reconstruction algorithms.
In cryo-EM, poses and conformations have real physical meanings.
Thus, we propose to measure the entanglement between poses and conformations as
$E_{entangle} = \mathcal{E}_{pose} + \mathcal{E}_{z}$,
where $\mathcal{E}_{pose}$ and $\mathcal{E}_{z}$ are 
pose and conformation label $z$ prediction errors, respectively.
Lower entanglement values indicate better disentanglement.
The pose error
$\mathcal{E}_{pose} = \mathcal{E}_{rot} + \mathcal{E}_{trans}$.
The generalized classification error $\mathcal{E}_{z}$ of multi-classification can be measured using the mean ($\mu_i$) and standard deviation ($\sigma_i$) of each class (Eq.~\ref{eq:E_entangle}).
$N_{pair}$ is the number of unique pairs. 

\begin{equation}
  \mathcal{E}_{z} =
    \frac{1}{N_{pair}}
      \sum_{i<j}
      \exp
      \left(-\frac{(\mu_i-\mu_j)^{2}}{\sigma_i\sigma_j}\right)
  \label{eq:E_entangle}
\end{equation}

\section{Experiments}
\subsection{Reconstructions on Simulated Datasets}
\subsubsection*{Experimental Setup}
To evaluate our method against ground truth values and other methods,
we created two simulated heterogeneous cryo-EM datasets
to quantitatively and qualitatively assess the performance
of \methname{} (more details see supplementary material).

In the first dataset (termed ``\textbf{80S-bimodal}'' hereafter), two ground-truth volumes were taken from a published
dataset~\cite{zhong2021cryodrgn, ellen_d_zhong_2021_4355284},
which correspond to the rotated and unrotated states of the 80S ribosome protein.
We created two simulated datasets with 20k/100k images
(image size $D=128$; 3.77 \AA{}/pixel) using
the cryo-EM image formation model, with an equal number
of projection images for these two ground-truth volumes.

Projection angles were sampled uniformly from the 3D rotation group $SO(3)$.
2D image translations were sampled from a uniform distribution between $[-8, 8)$ pixels along X and Y axes.
Gaussian noise was added to reach a signal-to-noise ratio (SNR) of -10 dB.
CTF parameters were taken from an experimental dataset EMPIAR-10028~\cite{wong2014cryo}.

The second dataset (termed ``\textbf{1D-motion}'') contains 100k images. It was generated from a 1-dimensional (1D) rigid-body rotational motion of a
protein fragment. We chose 10 representative states to generate the ground truth volumes.
Projection images (image size $D=64$; 3.0 \AA{}/pixel) were generated in the same way as the ``80S-bimodal'' dataset.

We trained \methname{} with the dimension of the latent variable $z$ set to 8 and activated the update of conformational states after 50 epochs.
The ADAM optimizer with learning rate of $1.5\times10^{-4}$ was used to optimize our model. The size of batches is set to 128 and 512 respectively for the two datasets.
cryoDRGN2~\cite{Zhong_2021_ICCV} was trained with its default parameters. The size of batches was set to 24 and 48 respectively for the two datasets.
cryoFIRE~\cite{Levy_2022_cryoFIRE} was also trained with its default settings. To better comparison with ours, we set the same size of training batches and activate the conformation MLP after the same epochs.
We trained all of the methods on a single NVIDIA V100 32GB.

For visualization, we use principle component  analysis (PCA) to transform conformational state $z_i$ into principle components (PC).
After convergence, we generate 3D volumes by taking the centroids of the clusters in the reduced principle component space and inverse-transform them into the corresponding $z_i$ values.

\subsubsection*{Benchmark on the 80S-bimodal datasets}
\label{sec:80s_benchmark}
With 80S-bimodal datasets, we compare \methname{} with cryoDRGN2~\cite{Zhong_2021_ICCV} and cryoFIRE~\cite{Levy_2022_cryoFIRE}, and evaluate the performance of heterogeneous reconstruction from the following metrics:

\begin{itemize}
    \item Rot.: the median squared error of rotation matrices using the Frobenius norm.
    \item Trans.: the median squared error of the predicted translation vectors (in pixels).
    \item Res.: the FSC resolution (in pixels) is the inverse of the spatial frequency corresponding to a Fourier Shell Correlation (FSC) coefficient of 0.5 between the predicted and ground-truth volumes.
    \item Err.: the percentage of mis-classified images over the total number of images
\end{itemize}

\begin{table}[t]
  \caption{Benchmark results for the 80S bimodal datasets.
    The best results are highlighted in bold. FSC resolutions
    are listed for both volumes.}
  \label{table:80S-bimodal}
  \begin{center}
  \begin{tabular}{lccc}
  \toprule
 Methods 	&	  cryoDRGN2 	&	cryoFIRE &  \methname{}	\\
    & (non-amortized) & (amortized) & (amortized) \\
 \midrule
 \textbf{20k dataset} \\

Rot. ($\downarrow$) 	&	\textbf{2e-4} 	&	 1.7 & 5e-4 	\\
Trans. ($\downarrow$) 	&	 8.7 	&	 10.9 & \textbf{1.4}	\\
Res. ($\downarrow$) 	&	 \textbf{2.14/2.14} 	&	4.82/4.04  & 	2.61/2.63 \\
Err. ($\downarrow$) 	&	 \textbf{0\%} 	&	38\% & 0.01\%	\\

\midrule

\textbf{100k dataset} \\

  Rot. ($\downarrow$) 	&	\textbf{2e-4} 	&	 5e-4 & 3e-4 	\\
  Trans. ($\downarrow$) 	&	 5.7 	&	 9.7 & \textbf{0.9}	\\
Res. ($\downarrow$) &	 \textbf{2.03/2.06} 	&	2.64/2.61  & 2.54/2.55	\\
Err. ($\downarrow$) 	&	 0.07\% 	&	0.02\% & \textbf{0.009\%}\\
    \bottomrule
    \end{tabular}
    \end{center}
\end{table}

\begin{figure}[t]
  \begin{center}
    \includegraphics[width=0.8\linewidth]{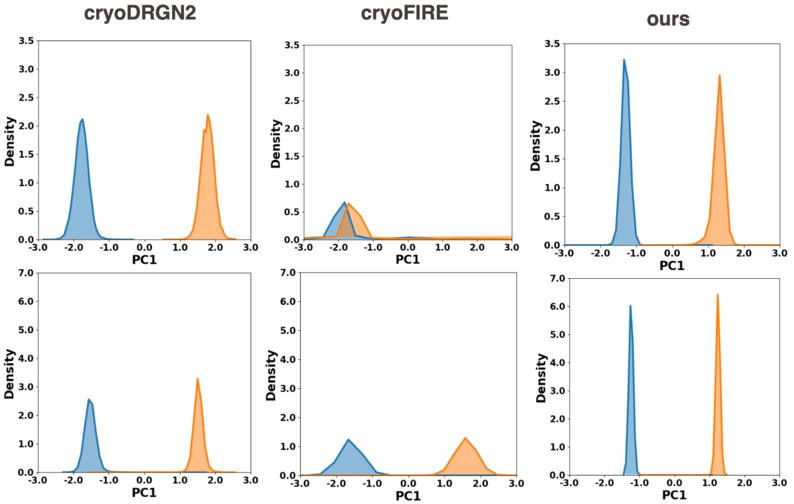}
  \end{center}
     \caption{Distributions of predicted conformations along the PC1 axis for the 20k (\textbf{top}) and 100k (\textbf{bottom}) 80S-bimodal datasets.}
  \label{fig:80s_bimodal}
\end{figure}

\begin{figure}[thb]
  \begin{center}
    \includegraphics[width=0.9\linewidth]{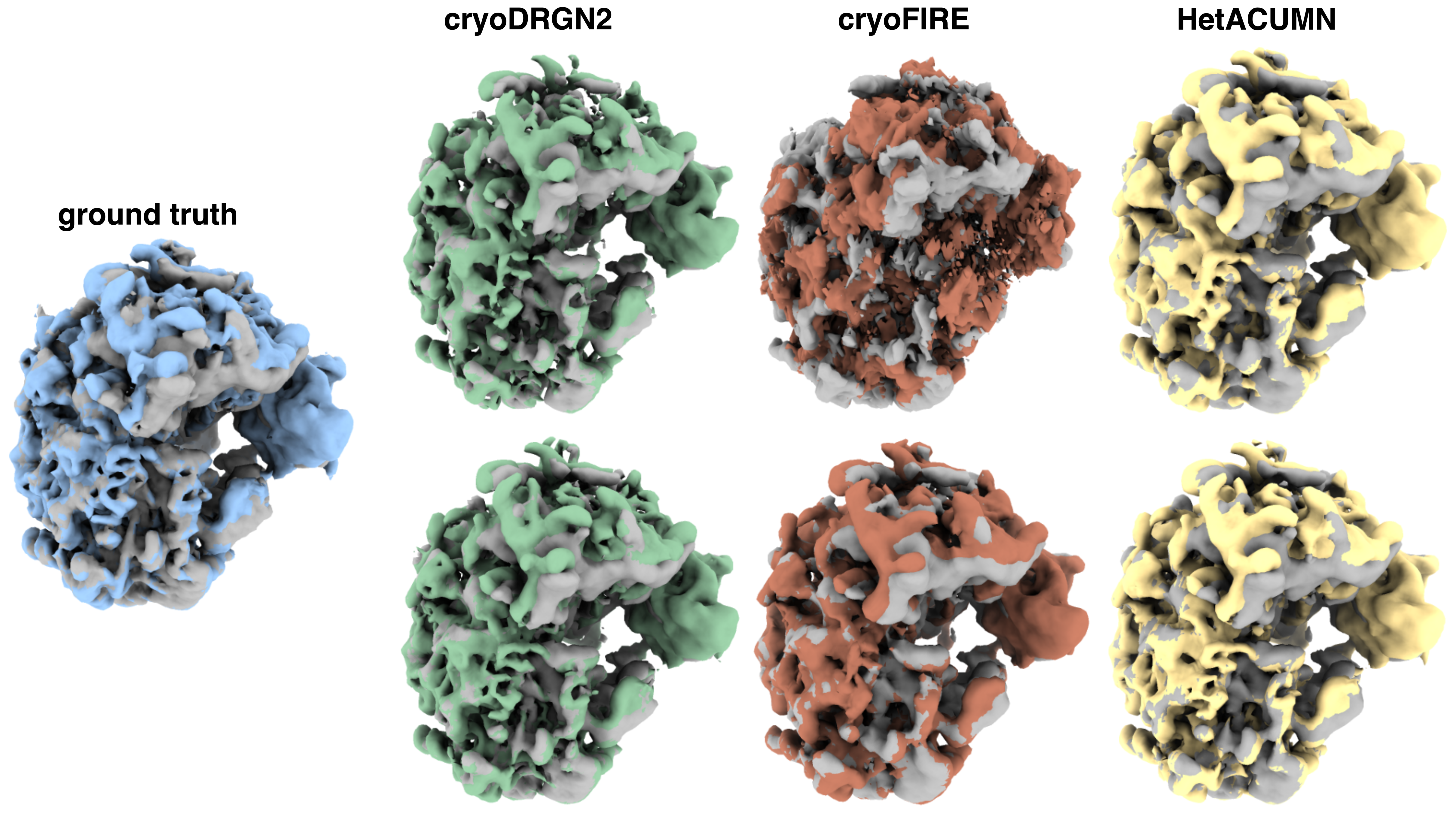}
  \end{center}
     \caption{Visualization of the ground-truth volumes and predicted volumes from cryoDRGN2, cryoFIRE, and \methname{}, reconstructed from  the 20k (\textbf{top}) and 100k (\textbf{bottom}) 80S-bimodal datasets. The two conformational states are superposed with one state shown in color and the other in grey.}
  \label{fig:80s_bimodal_vol}
\end{figure}

The performance of the three methods on the 80S bimodal datasets is presented in Table~\ref{table:80S-bimodal} as well as Figure~\ref{fig:80s_bimodal} and \ref{fig:80s_bimodal_vol}.
CryoDRGN2~\cite{Zhong_2021_ICCV}
performed well on both the 20k and 100k datasets thanks to
the less scalable but more accurate exhaustive pose search technique.
CryoFIRE, an amortized inference method, performed poorly on the 20k dataset but its performance
improved greatly when the dataset size increased to 100k.
\methname{}, also based on amortized inference, outperformed cryoFIRE
in all metrics, and achieved
comparable pose estimation accuracy as cryoDRGN2.

Among all metrics, the classification error is the most important one,
which indicates the ability of the model to distinguish different conformations.
For the small dataset (20k), cryoDRGN2 correctly classified all images.
CryoFIRE, on the other hand, mis-predicted the classification labels for
a large portion of the dataset (38\%).
\methname{} performed comparable to cryoDRGN2 and only mis-classified two images (0.01\%).
When the dataset size was increased to 100k, cryoDRGN2's classification error increased
to 0.07\%. CryoFIRE's performance on this larger dataset was slightly better than cryoDRGN2
(0.02\%). \methname{} outperformed both cryoDRGN2 and cryoFIRE and reduced the classification
error to 0.009\%. The poor performance of cryoFIRE on the smaller dataset (20k) indicates
a severe limitation of this method.
We argue that this is not an inherent drawback of amortized inference.  \methname{} is a better alternative when the data and/or computational resource is limited, which performed well on both small and large datasets.

Aside from these metrics, an interesting observation from Figure~\ref{fig:80s_bimodal} is that
the latent-$z$ space of \methname{} (after dimension reduction) showed much more
concentrated distributions than cryoDRGN2 and cryoFIRE. As the number of ground-truth
conformations increase, narrower $z$ value distributions around cluster centers become
much more beneficial (see 1D-motion dataset benchmark results below).

\subsubsection*{Benchmark on the 1D-motion dataset}
\begin{figure}[!t]
  \begin{center}
    \includegraphics[width=0.9\linewidth]{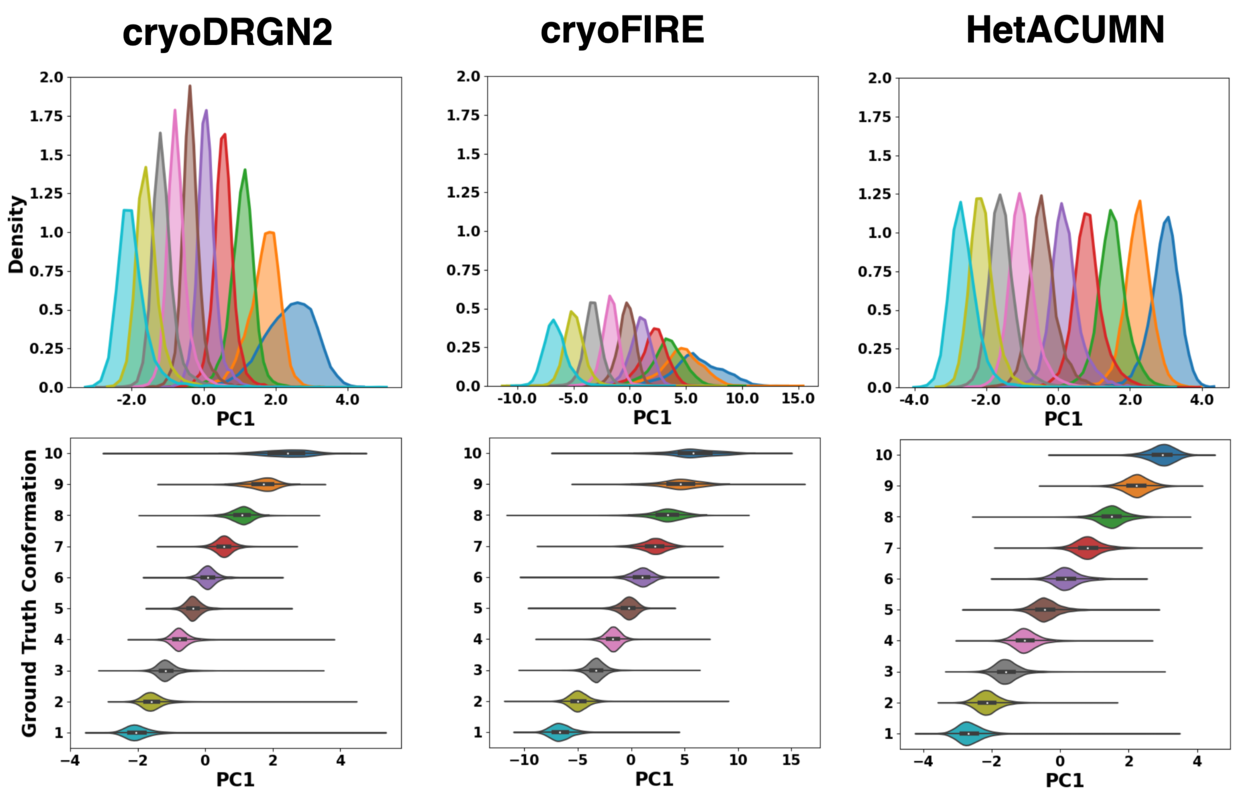}
  \end{center}
     \caption{Distribution of the latent-$z$ for benchmark tests on the 1D-motion dataset (100k).
     \textbf{(a)} probability density distribution of the latent-$z$ along the PC1 axis after dimension reduction.;
     \textbf{(b)} violin plots for the latent-$z$ statistics for each class.
     }
  \label{fig:7bcq_class10}
\end{figure}

The 1D-motion dataset consists of 10 distinct conformations, presenting a challenging task for conformation classification. The benchmark results are shown in Table~\ref{table:1D-motion} and Figure~\ref{fig:7bcq_class10}.
Besides the pose (rotation and translation) error,
we quantified the classification accuracy using the Spearman correlation coefficient (\textbf{Corr.}) between the ground-truth labels and the projections of latent-$z$ along the first principal component PC1.
\methname{} outperformed cryoDRGN2 and cryoFIRE in all metrics.
Since there are equal number of images for the 10 ground-truth classes,
we expect the latent-$z$ for each class to be similarly distributed around the probability density peak values.
A closer look revealed that these distributions
are quite uneven for cryoDRGN2 and cryoFIRE ((Figure~\ref{fig:7bcq_class10}). Especially for the last several classes, the distribution estimated by cryoFIRE become some degenerate peaks that cannot be distinguish between them. In contrast,
the shapes of the latent-$z$ distribution for \methname{} are similar and distinguishable regardless of the underlying conformation class.

\begin{figure}[!t]
  \begin{center}
    \includegraphics[width=0.8\linewidth]{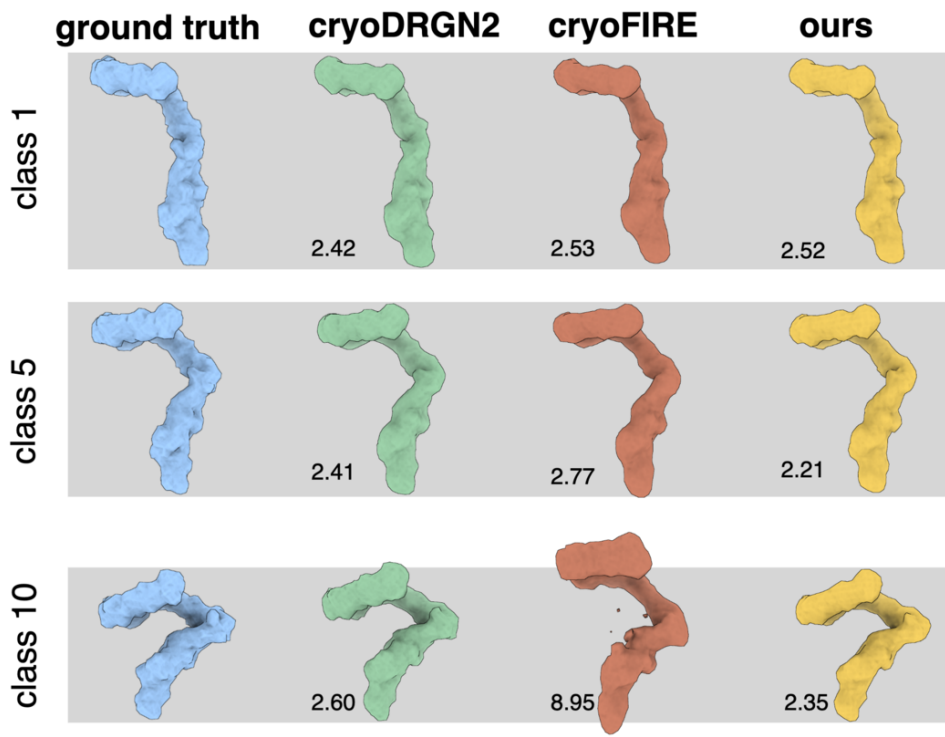}
  \end{center}
     \caption{Visualization of the ground-truth and predicted volumes for the 1st, 5th, and 10th
     conformational classes, reconstructed from the 1D-motion dataset (100k). FSC resolutions (in pixels) are shown at the lower left corners.
     }
  \label{fig:7bcq_class10_vol}
\end{figure}

In addition, we generated visualization of the volumes of the 1st, 5th, and 10th classes from cryoDRGN2, \methname{}, and the groundtruth (see Figure~\ref{fig:7bcq_class10_vol}). Comparing with the ground truth, it is no obvious difference for cryoDRGN2 and cryoFIRE on the first two volumes. But for the 10th class, there is some visible inconsistency of predicted volume against the ground truth. This observation is consistent with the latent-$z$ distribution mentioned above.

\begin{table}[htbp]
  \caption{Benchmark results for the 1D-motion dataset (100k). }
  \label{table:1D-motion}
  \begin{center}
  \begin{tabular}{lccc}
  \toprule
 Method 	&	  cryoDRGN2 	&	cryoFIRE &  \methname{}	\\
 \midrule
Rot. ($\downarrow$) 	&	8e-3 	&	 2e-2 & \textbf{7e-3} 	\\
Trans. ($\downarrow$) 	&	 4.3 	&	 5.4 & \textbf{3.1}	\\
Corr. ($\uparrow$)	&	 0.963 	&0.951 & \textbf{0.976}	\\
  \bottomrule
  \end{tabular}
  \end{center}
\end{table}

\subsubsection*{Latent-space Entanglement}
Using the metric defined in Eq.~\ref{eq:E_entangle},
here we quantitatively compare the latent-space entanglement of 
cryoDRGN2, cryoFIRE, and \methname{}.
For all three datasets (bimodal 20k, bimodal 100k, 1D motion),
\methname{} showed the least entanglement of the latent space for
pose and conformation, and outperformed cryoFIRE by a significant amount.
It is worth noting that cryoDRGN2 estimates the poses through 
search-based algorithms rather than using a neural network.
This design makes cryoDRGN2 less prone to latent-space entanglement than cryoFIRE.
The fact that \methname{} performed still better than cryoDRGN2 demonstrates the 
value of adding the conditional pose prediction task during training.

\begin{table}[htbp]
    \vspace{-5pt}
    \caption{Entanglement $E_{entangle}$ ($\downarrow$) for all three methods.}
    \label{table:entanglement}
    \vspace{-1pt}
    \begin{center}
    \begin{tabular}{lccc}
    \toprule
   Methods 	&	  cryoDRGN2 	&	cryoFIRE &  \methname{}	\\
   \midrule
  bimodal 20k &	7.3e-4	&	2.08 & \textbf{6.8e-4}\\
  bimodal 100k &	1.2e-3 	& 1.3e-3	& \textbf{4.4e-4}\\
  1D motion 	&	4.6e-2	&	8.2e-2 & \textbf{1.0e-2}	\\
  \bottomrule
  \end{tabular}
  \end{center}
  \vspace{-5pt}
  \end{table}

\subsection{Reconstructions on an Experimental Dataset}
\subsubsection*{Experimental Setup}

To test the applicability of \methname{} on real
experimental cryo-EM data, we chose
the precatalytic spliceosome dataset from the
EMPIAR database (ID: EMPIAR-10180)
~\cite{Plaschka_2017_5nrl,5nrl_empiar_url},
which was downsampled to an image size of $D=128$ (4.2475 \AA{}/pixel).
We run \methname{} with the same training settings as for the simulated datasets.

\subsubsection*{Performance}
\begin{figure}[htbp]
  \includegraphics[width=0.9\linewidth]{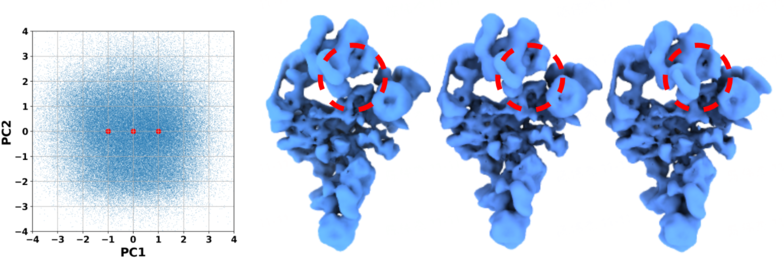}
   \caption{Visualization of the latent conformation space and volumes predicted by \methname{} using the spliceosome dataset (EMPIAR-10180) from real experiments. Three representative volumes are shown here, sampled along the PC1 axis (PC1 values from left to right: -1, 0, 1)}.
\label{fig:spliceosome}
\end{figure}

For experimental datasets, the distribution of noise is often more complex than simulated datasets. Based on our simplified noise estimation method (Equation~\ref{eq:snr}), the SNR of spliceosome datasets is around $-11.7$ dB, which is worse than the simulated datasets above.
We predicted the latent $z$ for the full dataset, then sampled three points along the first component axis and generated the corresponding volumes. As shown in Figure~\ref{fig:spliceosome}, changes in the shape of the reconstructed volumes indicate a large flexible motion in the continuous conformational space. The corresponding predicted volumes and the distribution of predicted latent $z$ along the first two principal component axes are similar to that from the literature~\cite{zhong2021cryodrgn}.

\section{Ablation Study}
\label{sec:ablation}

\begin{figure}[!t]
  \begin{center}
    \includegraphics[width=0.9\linewidth]{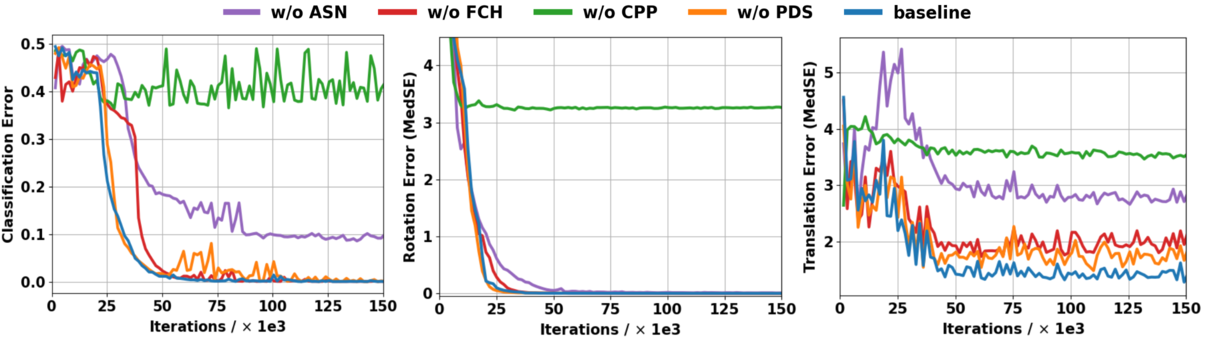}
  \end{center}
     \caption{Ablation study for \methname{} benchmarked on the 80S-bimodal dataset (20k).}
  \label{fig:ablation}
\end{figure}

As mentioned in section \ref{sec:method}, we proposed a novel training task \textbf{CPP} to explicitly enforce the disentanglement of conformation state and pose. Besides, \methname{} also uses three strategies to improve the performance of pose estimations and conformation classifications.
\begin{itemize}
  \item Frozen Conformation-prediction Head (\textbf{FCH}): Both cryoFIRE~\cite{Levy_2022_cryoFIRE} and \methname{} freeze the conformation-prediction head during the early training stage in order to mitigate the difficulties of the volume generation and pose estimation.
  \item Posterior $z$-Distribution Sampling (\textbf{PDS}): To generate synthetic images, we sampled conformational states $z$ from the posterior distribution $q(z|\mathcal{I})$ instead of the prior distribution $p(z)$ (i.e., the standard Gaussian distribution).
  \item Adaptive Synthetic Noise (\textbf{ASN}): in the CPP task, we generated adaptive noise using a fixed SNR instead of a fixed noise variance.
\end{itemize}
To quantify the necessity of the CPP task as well as the three strategies mentioned above, we designed four ablation experiments using the 80S-bimodal (20k) dataset, while keeping other training setup unchanged.
\methname{} with a full setup was used as the baseline.

As shown in Figure~\ref{fig:ablation}, without the CPP task, large errors in conformation classifications and pose estimations are observed. One explanation is that conformational states and poses are both latent variables without direct supervision. It is difficult to  learn their corresponding physical meaning through indirect supervision with an image reconstruction loss especially during the early training phase. This difficulty has also been observed in other studies~\cite{DBLP:journals/corr/abs-2106-14108}.
The CPP task adds direct supervision to the pose variable, making it disentangled from the conformational state variable.

In addition to the CPP task, other strategies also contribute to reducing
estimation errors (Figure~\ref{fig:ablation}).
For image translation estimation, all three strategies (FCH, PDS, and ASN) are necessary. For conformation classification, ASN appears to be critical. During the early training phase, without ASN, extremely strong noise can overwhelm the signal of images and mislead the encoder.
Both FCH and PDS helped to reduce the estimation instabilities.
For rotation (orientation) estimation, the absence of ASN and FCH slightly delayed the
convergence while removing PDS seemed to have little effect.
Therefore, we recommend using the full setup when employing \methname{} for heterogeneous cryo-EM reconstruction tasks.

\section{Conclusions}

In this paper, we discuss the challenges of heterogeneous cryo-EM reconstruction, including low signal-to-noise ratio, unknown poses, and complex distributions of underlying 3D structures.

To address these challenges, we propose a self-supervised deep learning framework called \methname{} based on amortized inference. By alternating the variational image reconstruction task and the conditional pose prediction task, the VAE-based architecture explicitly enforces the disentanglement of the conformation and pose latent space. The experiments on simulated datasets show that \methname{} outerformed other armotized-inference-base methods like cryoFIRE. On the other hand, our method has comparable accuracy of pose estimation and even better performance in estimating conformational distribution than non-amortized methods. Furthermore, we demonstrated that \methname{} can also be used on experimental datasets.

There are still some limitations of our method. For exmaple,
the latent $z$ space is capable of representing significant amounts of conformational variations. However, real experimental datasets is often  complex and the reduced $z$ space through principal component analysis may not be the best choice for conformational classification. Further study in this direction is needed.

Overall, we believe there is a great potential in applying deep learning approaches for heterogeneous cryo-EM reconstruction. It provides a new path for improving the accuracy and efficiency of this challenging task.

{\small
\bibliographystyle{ieee_fullname}
\bibliography{refs-hetem.bib}
}

\clearpage

\setcounter{page}{1}
\setlength{\itemsep}{0pt}

\renewcommand\thesection{\Alph{section}}
\renewcommand\thesubsection{\thesection\arabic{subsection}}
\renewcommand\thefigure{S\arabic{figure}}
\setcounter{figure}{0}
\setcounter{section}{0}


\section{Model architecture}
\subsection{Encoder}
\begin{figure}[htbp]
  \begin{center}
  \includegraphics[width=0.8\linewidth]{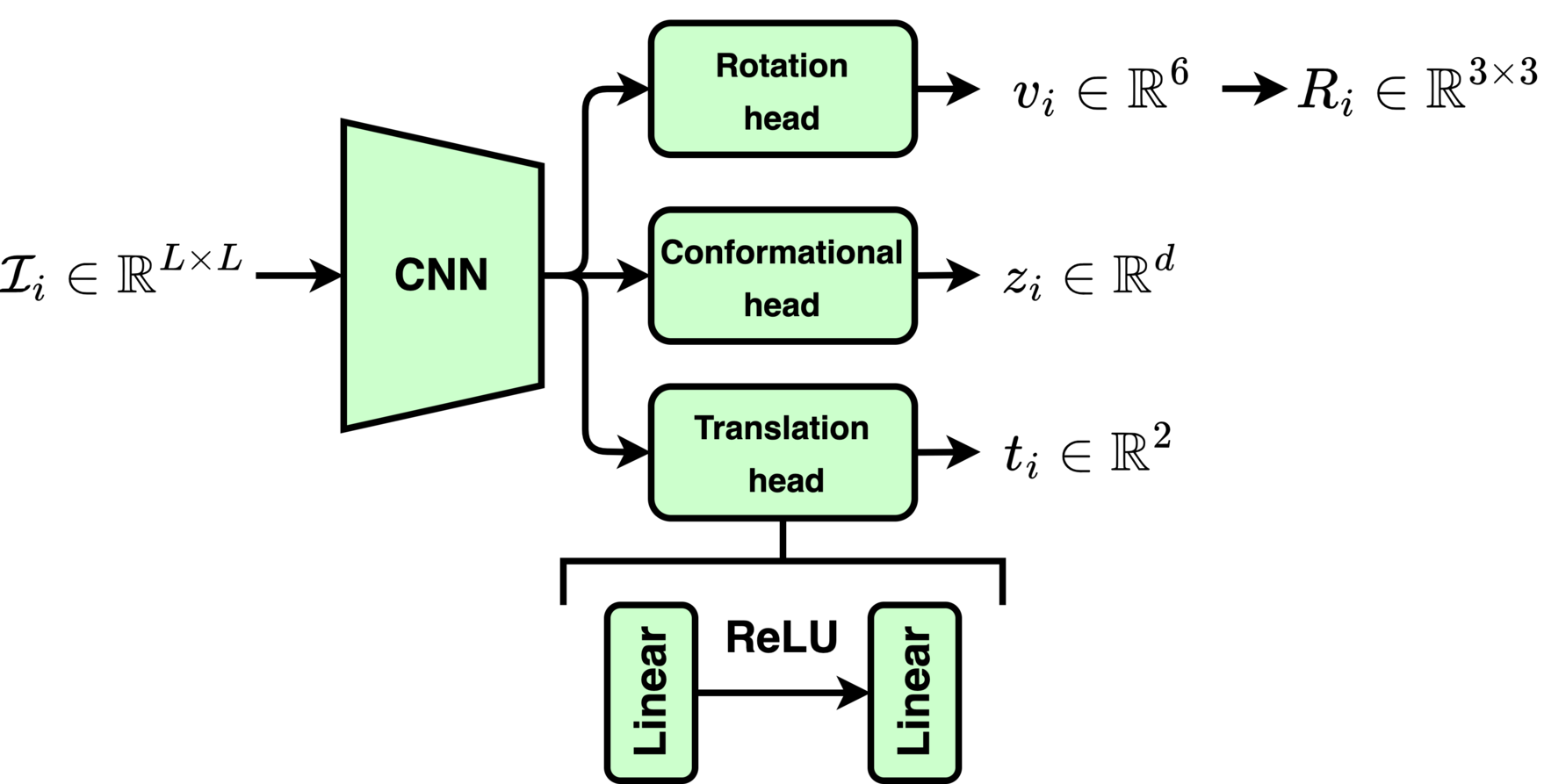}
  \end{center}
   \caption{Architecture of the encoder.}
\label{fig:encoder}
\end{figure}
Figure~\ref{fig:encoder} shows the architecture of the \methname{} encoder. The architecture of Convolutional Neural Network (CNN) is the same as the standard ResNet-18~\cite{he2016deep} without the full connection layer. The output of CNN is flattened and fed into three parallel fully connected neural networks with a hidden layer dimension of 256.

\subsection{Decoder}
\begin{figure}[htbp]
  \begin{center}
  \includegraphics[width=0.8\linewidth]{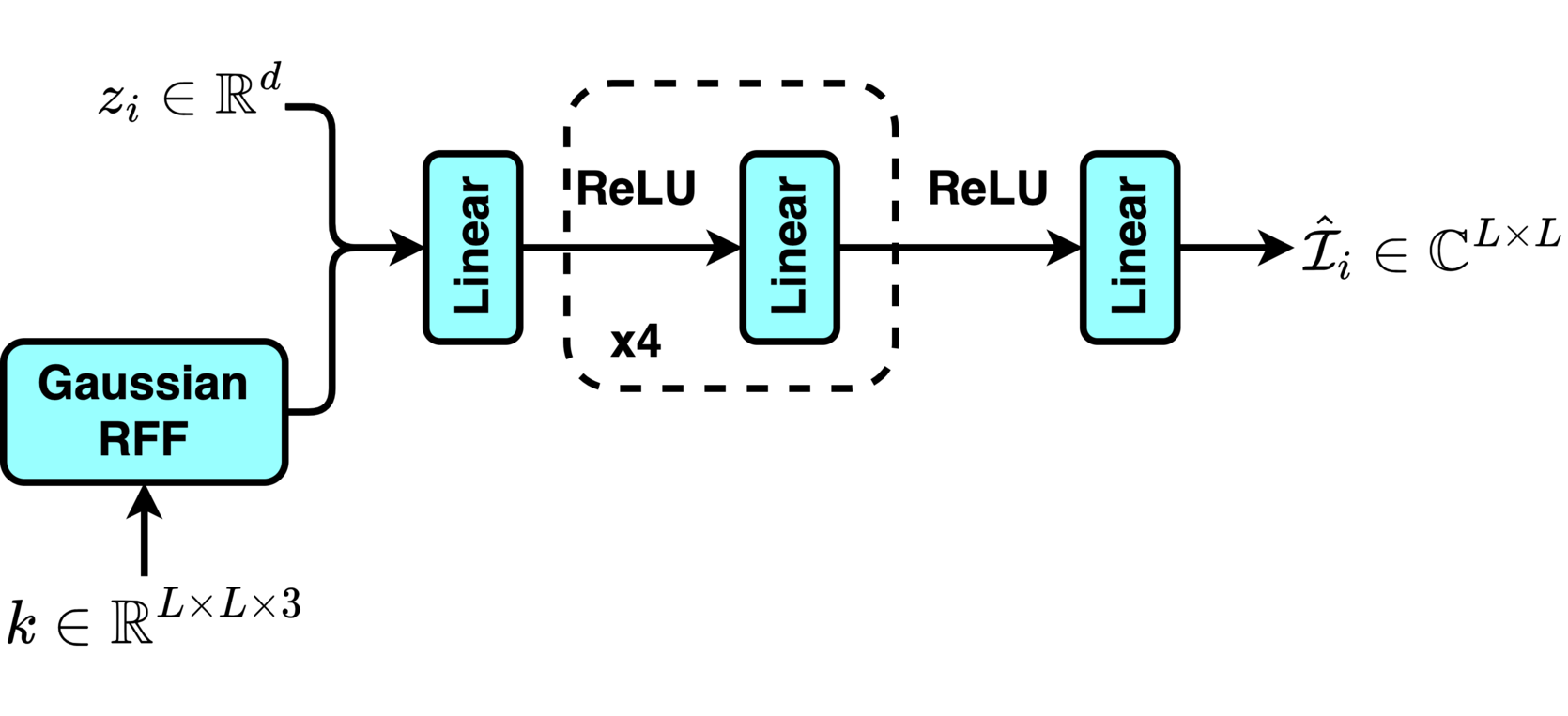}
\end{center}
   \caption{Architecture of the decoder.}
\label{fig:decoder}
\end{figure}
Figure~\ref{fig:decoder} shows the architecture of the \methname{} decoder. The Fourier-space coordinates $k\in\mathbb{R}^{L\times L\times3}$ are encoded by the Gaussian random Fourier Features (RFF) mapping~\cite{rahimi2007random, tancik2020fourier}. Then, the conformational state $z_i$ and encoded coordinates are fed into the coordinate-based fully connected neural networks with a hidden layer dimension of 256.

\section{Dataset preparation}
\label{SI:dataset_prep}

\subsection*{Simulated datasets}
The simulated datasets were created from ground-truth volumes
using the forward cryo-EM image formation model as described in
the main text.
Each ground-truth volume was first rotated
to a certain orientation sampled uniformly
from the 3D rotation group SO(3).
A projection image was formed by integrating the voxel values
along z axis (orthogonal to the projection plane).
CTF effects were added to the projection image with
CTF parameters uniformly from the experimental dataset
EMPIAR-10028.
Then, Gaussian noise was added to the projection image to reach an SNR of -10 dB.
Finally, a random 2D image translation was added which was sampled from a uniform distribution within the range of $[-8, 8]$ pixels.

For the ``80S-bimodal'' dataset, the two ground-truth volumes
(rotated and un-rotated states of the 80S ribosome protein)
were selected from a previous published dataset~\cite{zhong2021cryodrgn,ellen_d_zhong_2021_4355284}.
Each volume has a cubic shape with edge length $D=128$.
The generated projection image has a pixel size of 3.77 \AA{}/pixel.
As for the ``1D-motion'' dataset,
the protein fragment (residues 177--330 from chain A with hydrogen atoms removed)
was taken from a published atomic structure of the
ASCT2 protein (PDB ID: 7BCQ)~\cite{Garibsingh_2021_7bcq}.
The 1D rigid-body motion was generated by rotating atoms
around the backbone C-CA bond of residue GLY249 (see Figure~\ref{fig:10class}).
The rotation angles were taken from the following sequence
$\{9^{\circ}, 18^{\circ},\dots,90^{\circ}\}$.

\subsection*{Experimental dataset}
The precatalytic spliceosome experimental dataset was downloaded
from the EMPIAR database (ID: EMPIAR-10180)
~\cite{5nrl_empiar_url,Plaschka_2017_5nrl}.
The original dataset was filtered to a final size of
139,722 images using the indices from a previous study
~\cite{ellen_d_zhong_2021_4355284} and downsampled to an image size
of $D=128$ and a pixel size of 4.2475 \AA{}/pixel.
Since the raw images were significantly off-centered, all images were recentered using
the published image translation values~\cite{Plaschka_2017_5nrl}.

\section{Performance with time constraints}
\begin{figure*}[htbp]
    \begin{center}
    \includegraphics[width=0.8\textwidth]{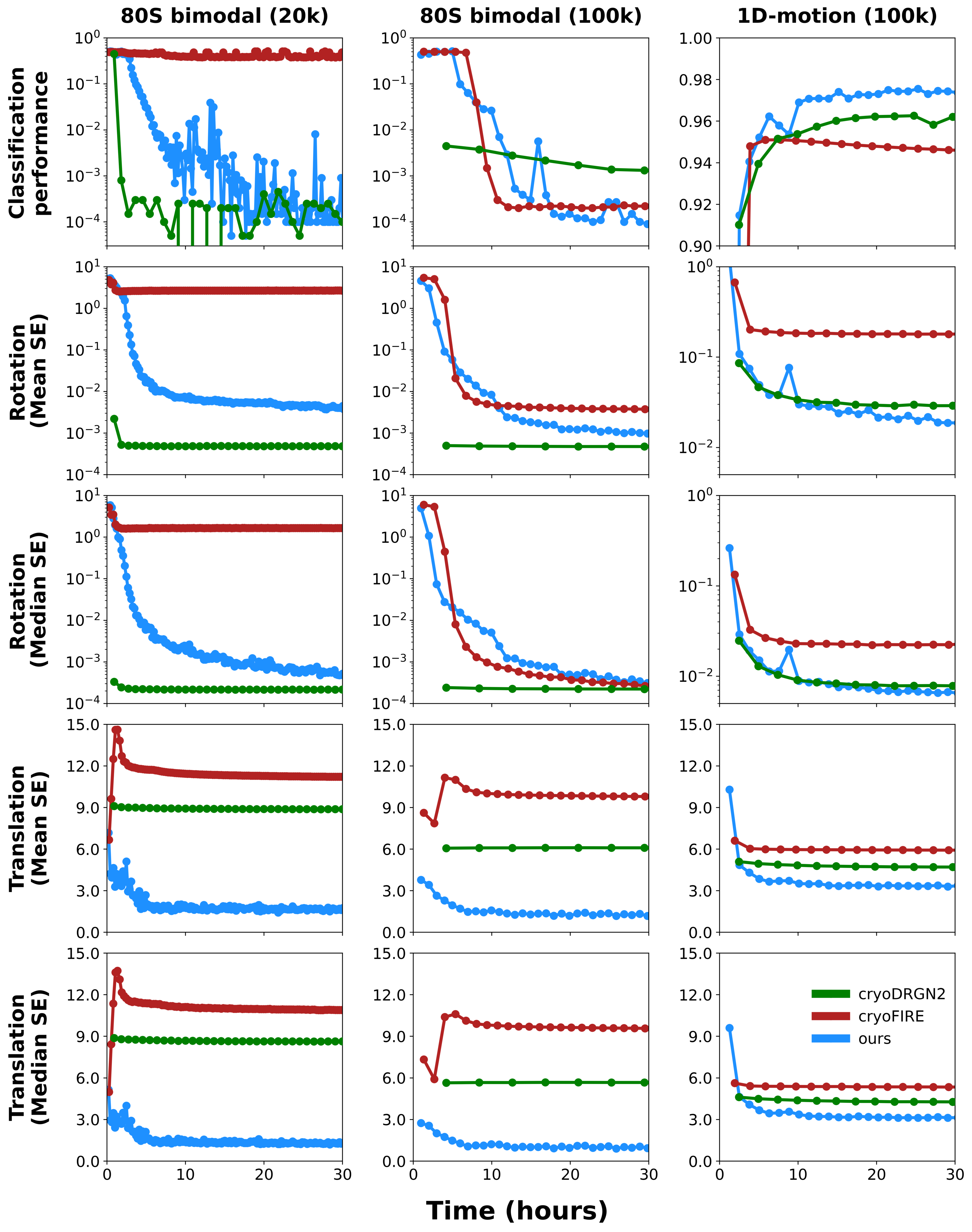}
    \end{center}
     \caption{
        The performance of cryoDRGN2 ({\textcolor{limegreen}{\textbf{green}}}),
        cryoFIRE ({\textcolor{red}{\textbf{red}}}),
        and \methname{} ({\textcolor{blue}{\textbf{blue}}})
        within a wall-clock time of 30 hours,
        which were benchmarked on three datasets (left to right):
        80S bimodal (20k),
        80S bimodal (100k),
        1D-motion (100k).
        The classification performance
        was quantified in terms of confusion error ($\downarrow$)
        for the two 80S bimodal datasets
        or Spearman correlation coefficient ($\uparrow$)
        for the 1D-motion dataset.
        The rotation error ($\downarrow$) is defined as the
        Frobenius norm of the difference matrix between the ground-truth
        and predicted rotation matrices.
        The translation error ($\downarrow$)
        is defined as the L2 norm of the difference vector (in pixels)
        between the ground-truth and predicted translation vectors.
        Both mean and median squared errors (SE) for rotation and translation
        are reported here.
     }
  \label{fig:convergence}
  \end{figure*}

For real-world applications, cryo-EM data analyses
are often subject to time constraints,
especially in industrial settings.
Therefore, it is important to consider the performance
within a given wall-clock time.
Here we show the performance of cryoDRGN2, cryoFIRE, and \methname{}
within a wall-clock time of 30 hours (Figure~\ref{fig:convergence}).
For all three datasets, \methname{} achieved similar or better performance
compared to cryoDRGN2 and cryoFIRE in terms of classification performance,
rotation error (mean and median),
and translation error (mean and median).
Note that Table 1 and 2 in the main text report the best performance
after reaching convergence in classification performance, not necessarily
at the end of training shown here.

\section{FSC resolution anslysis}
To quantify the volume reconstruction quality of the three methods,
we performed
Fourier shell correlation (FSC) analysis for all predicted
volumes against the ground-truth volumes.
Each volume was generated based on a representative classification latent vector $z_i^*$
for images of the corresponding ground-truth conformational class $i$.
To calculate $z_i^*$, we first performed PCA dimension reduction for all $z$ vectors
of the same ground-truth conformational class $i$.
Then, we projected these vectors along the first principal component (PC1)
and selected the $z$ corresponding to the highest probability density
of this 1D $z$-distribution (see Figure 3 and 5 in the main text).
Following the convention in the field of cryo-EM,
the resolutions of the predicted volumes were estimated at an FSC cutoff of 0.5,
i.e., the spatial frequency at which the FSC curve drops to 0.5.
As a common practice, the resolutions reported here are
expressed in terms of the inverse of the spatial frequencies.
As a result, the lower the value, the better the resolution.

\begin{figure}[htbp]
    \begin{center}
    \begin{subfigure}{0.6\textwidth}
      \centering
      \includegraphics[width=1.0\linewidth]{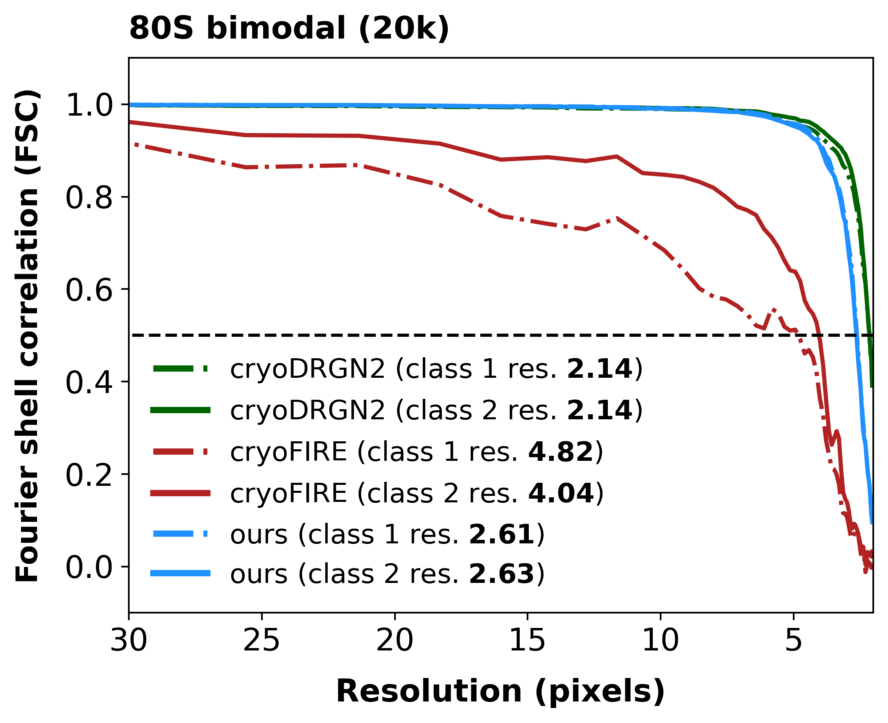}
      \vspace{10pt}
    \end{subfigure}

    \begin{subfigure}{0.6\textwidth}
      \centering
      \includegraphics[width=1.0\linewidth]{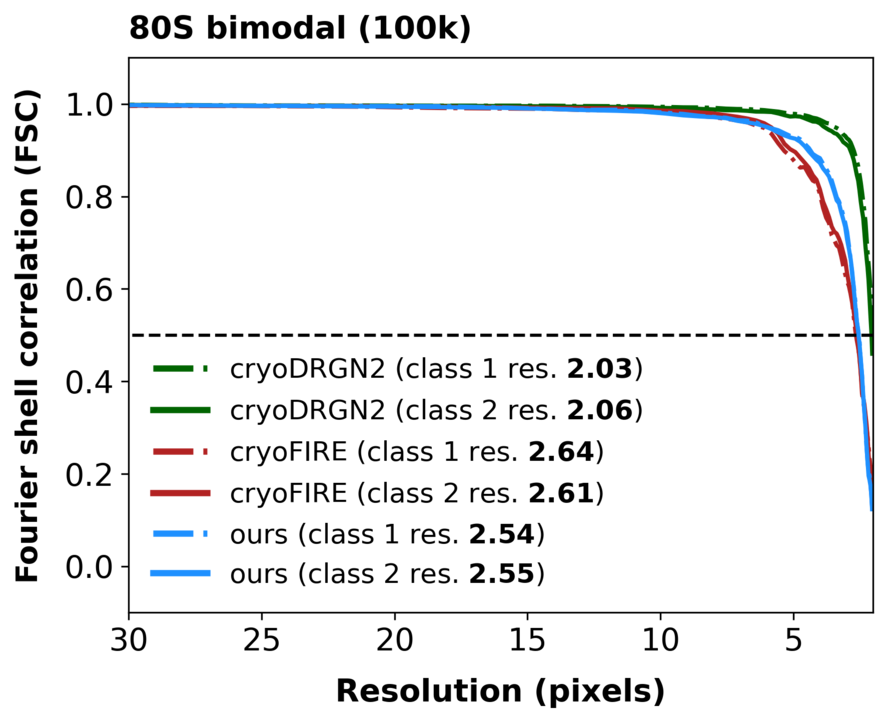}
    \end{subfigure}

    \end{center}
     \caption{Fourier shell correlation (FSC) analysis for the predicted
     vs.\ ground-truth volumes of both conformational classes from the 20k
     (\textbf{top})
     and 100k (\textbf{bottom}) 80S bimodal datasets.
     Resolutions (res.) in pixels of the predicted volumes are estimated at
     an FSC cutoff of 0.5 (dashed line) and marked in \textbf{bold} text.
     }
  \label{fig:80S_bimodal_FSC}
  \end{figure}

Figure~\ref{fig:80S_bimodal_FSC} shows the FSC
analysis for the two 80S bimodal datasets.
When using the 80S bimodal 20k dataset, the resolution of the reconstructed volumes
from cryoDRGN2 and \methname{} are similar.
However, the volume reconstructed by cryoFIRE showed much lower resolutions.
When using the 100k 80S bimodal dataset, all three methods achieved similar
reconstruction quality.

\begin{figure*}[htbp]
    \begin{center}
    \includegraphics[width=0.5\textwidth]{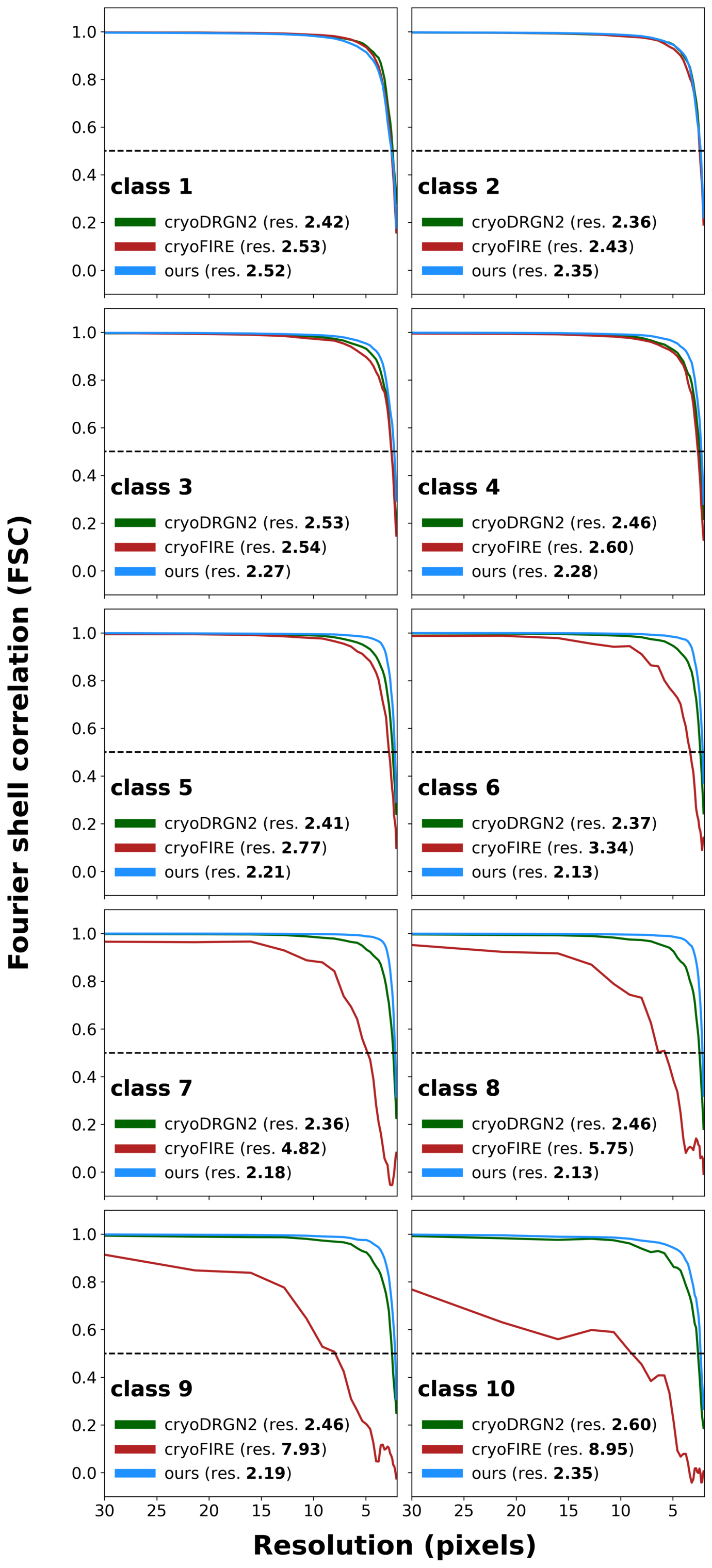}
    \end{center}
     \caption{Fourier shell correlation (FSC) analysis for the
     predicted vs.\ ground-truth volumes of
     each conformational class from the 1D-motion dataset.
     Resolutions (res.) in pixels of the predicted volumes are estimated at
     the FSC cutoff of 0.5 (dashed line) and marked in \textbf{bold} text.
     }
  \label{fig:1D_motion_FSC}
  \end{figure*}

\begin{figure}[htbp]
    \begin{center}
    \includegraphics[width=0.6\textwidth]{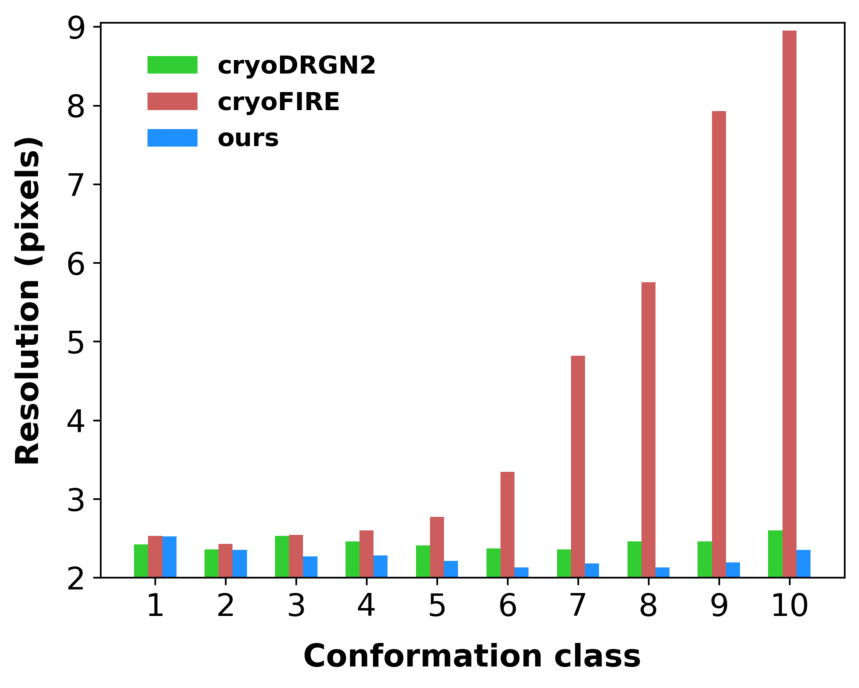}
    \end{center}
     \caption{Distribution of the FSC resolutions in pixels
     (a lower value means better resolution)
     for volumes from all 10 conformational classes of the 1D-motion dataset
      generated by cryoDRGN2, cryoFIRE, and \methname{}.
      }
  \label{fig:1D_motion_res_hist}
  \end{figure}

\begin{figure*}[htbp]
      \begin{center}
      \includegraphics[width=0.5\textwidth]{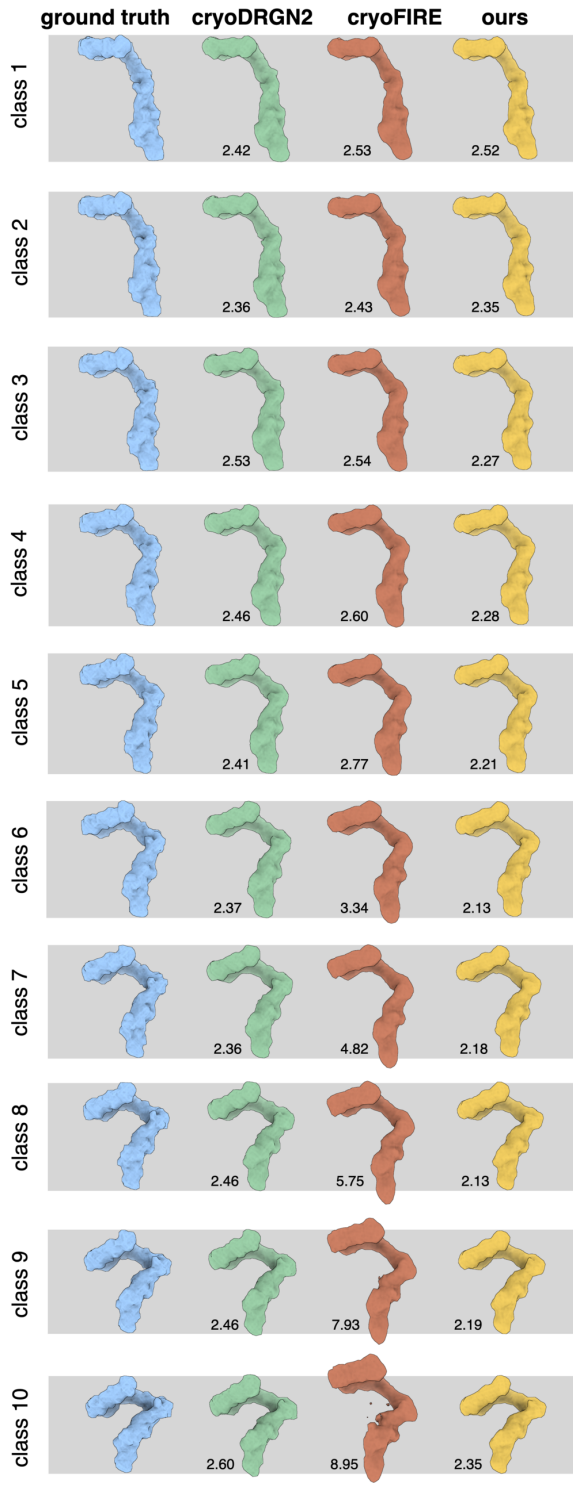}
      \end{center}
       \caption{Visualization of the ground-truth and predicted volumes of
       the 10 conformational classes in the 1D-motion dataset.
       FSC resolutions (in pixels) are shown at the lower left corners.
       The gray boxes serve as visualization aids for easier comparison.
       Their heights are adjusted to precisely fit the ground-truth volumes.
       \textbf{Top} to \textbf{bottom}: conformational class 1 to 10.
       \textbf{Left} to \textbf{right}: ground truth, cryoDRGN2, cryoFIRE, and \methname{} (ours).
       }
      \label{fig:10class}
  \end{figure*}

For the 1D-motion dataset,
\methname{} achieved better resolutions compared to cryoDRGN2
for all 10 conformational classes except for class 1.
On the other hand,
the resolutions of the volumes reconstructed by cryoFIRE were lower than
both cryoDRGN2 and \methname{}, and gradually deteriorated
from class 1 to class 10 (Figure~\ref{fig:1D_motion_FSC}).
For easier cross-comparison, we also plotted the distribution of the FSC resolutions
for all 10 conformation classes (Figure~\ref{fig:1D_motion_res_hist}).
Lastly, we show the visualization of the reconstructed volumes for all
10 conformation classes (Figure~\ref{fig:10class}).

\section{Comparison on experimental datasets}
\begin{figure}[htbp]
    \centering
    \includegraphics[width=0.8\linewidth]{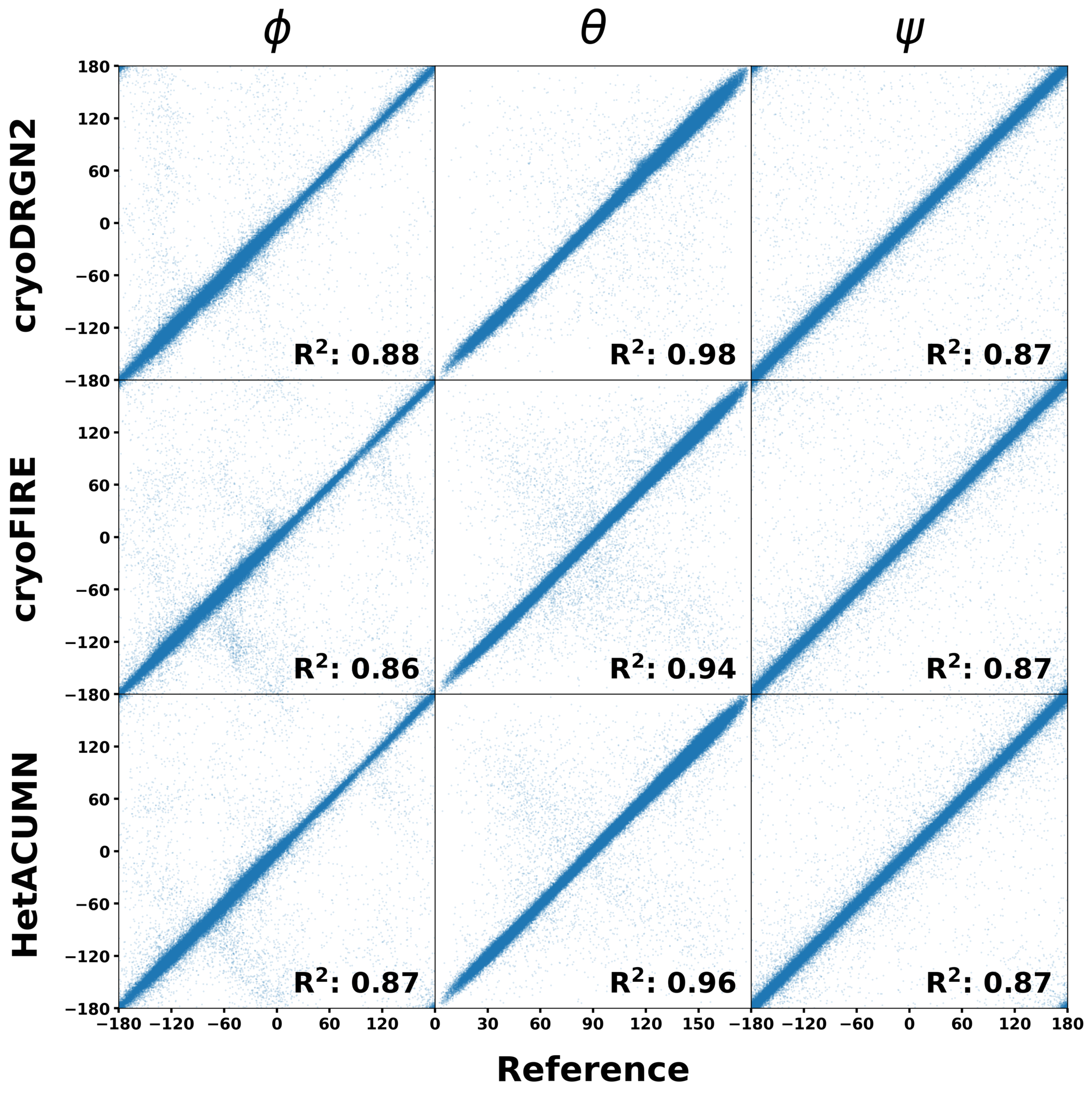}
    \caption{Comparison of the Euler angles between the published reference and the predictions for the experimental spliceosome dataset.
    $R^2$ values are shown in bold.}
    \label{fig:expt_euler}
  \end{figure}

The performance of \methname{} is comparable to cryoDRGN2 and cryoFIRE on the experimental dataset (Figure~\ref{fig:expt_euler}). Note that a precise cross-method comparison cannot be done because of the lack of ground-truth labels in real-world cryo-EM experiments.

\end{document}